# Interfacial spin glass mediated unusual spontaneous exchange bias effect in epitaxial self-assembled $La_{0.7}Sr_{0.3}MnO_3$:NiO nanocomposite thin films

Gyanendra Panchal,[1] R. J. Choudhary[1]*, Manish Kumar,[2] and D. M. Phase[1]

[1] UGC DAE Consortium for Scientific Research, University Campus, Khandwa Road, Indore 452001, India

[2] Pohang Accelerator Laboratory, POSTECH, Pohang 37673, South Korea

**Abstract:**

Zero field cooled spontaneous exchange bias (SEB) is observed in epitaxial $La_{0.7}Sr_{0.3}MnO_3$ (LSMO):NiO self-assembled nanocomposite thin films grown on (001) $SrTiO_3$ single crystal substrate by pulsed laser deposition. SEB is displayed by the novel asymmetry in the hysteresis loop of the composite film along with the field cooled conventional exchange bias (CEB) effect. The training effect shows that exchange bias relaxation is disorder mediated. It is revealed from DC magnetization results that such nanocomposite film divulges spin glass like behaviour, which is arising due to competing ferromagnetic and antiferromagnetic inteactions at the assorted interfaces of ferromagnetic LSMO and antiferromagnetic NiO. Also corroborated by x-ray magnetic circular dichroism (XMCD) measurements, we conclude that SEB is mainly originating due to ferromagnetic coupling of unstable interfacial antiferromagnetic spin due to NiO with the ferromagnetic LSMO at the disordered interface. These results of self-assembled thin films provide a useful input to realize and understand microscopic origin of SEB for device application.

*Corresponding author: ram@csr.res.in

**Introduction:**

Hetero-interfaces composed of magnetically active transition metal oxides exhibit very interesting phenomena such as interfacial ferromagnetism,[1] anomalous hall-effect,[2,3] interfacial superconductivity [4] in conventional horizontal interfacial layered geometry in the form of bilayer, multilayer and superlattices.[5–8] In the recent years, epitaxial self-assembled nanocomposite thin films are found to provide a new playground to enhance the physical properties and yield new functionalities via interplay among charge, orbital and spin degrees of freedom.[9,10,11] Synthesising of self-assembled two phase vertically aligned nanocomposite (VAN) film to create new device architecture is a rather new concept to enhance and tune electronic and magnetic properties via vertical strain tuning. Such architects have huge prospects for spintronic devices with certain advantage over conventional interfacial structure such as bilayers and superlattices.[12–15]

One of the most prominent interfacial phenomena which has been vigorously studied over last couple of decades is the exchange bias (EB) effect, arising from ferromagnetic (FM)-antiferromagnetic (AFM) interfacial magnetic coupling and the pinning effect at the heterointerfaces, which is well exploited in magnetic storage devices. Exchange bias is still a crucial factor for next generation magnetic storage device and its spintronics application.[16] In general EB effect is observed when FM/AFM interface is field cooled through Neel temperature ($T_N$) of AFM layer and the hysteresis loop is shifted along the field axis generally in the opposite direction to the cooling field. In conventional EB (CEB), the $T_N$ of AFM layer is below the Curie temperature ($T_C$). However in last few years unusual exchange bias is reported in variety of systems where the AFM layer is not required in the heterostructure, such as FM layer integrated with Pauli paramagnetic $LaNiO_3$,[17] nonmagnetic MgO,[18] multiferroic $BiFeO_3$,[19] and spin glass CuMn.[20] In these types of systems, many new fascinating exchange bias phenomena have emerged viz. positive exchange bias where the shift in the field cooled magnetic hysteresis loop is in the same direction as the applied biasing field;[21] unconventional exchange bias where the $T_C$ of FM layer is lower than the $T_N$ of AFM layer[22] and spontaneous exchange bias (SEB) where the unidirectional exchange anisotropy at the interface is created without field cooling the system, causing EB effect in zero field cooled condition also.[23]

Generally G-type (AFM) is not expected to pin the FM spin by exchange coupling because of compensated spin arrangement at the FM/AFM interface. However, recently unusual exchange bias was observed in epitaxial heterostructures comprising of FM-$La_{0.7}Sr_{0.3}MnO_3$ and G-type AFM-$SrMnO_3$, which was hugely correlated with interface spin frustration and attributed to the Dzyaloshinskii Moriya interaction.[24,25] The interplay of different exchange interactions at and across the interfaces plays a crucial role in establishing the EB effect in such FM and G-AFM heterostructure. Keeping this input in consideration, we aimed at manoeuvring such interplay of exchange interactions by enhancing the effective interfacial area between the two types of magnetic materials in the form of VAN structure. In the present study we observe that the competing magnetic exchange interactions at the extended interfacial region of epitaxial self-assembled $La_{0.7}Sr_{0.3}MnO_3$-NiO nanocomposite thin films lead to spin glass behaviour, which triggers spontaneous and unconventional exchange bias effect. The EB effect persists up to the spin glassy nature of the interface, with its demise EB effect also vanishes.

**Experimental:**

Self-assembled nano composite thin films of $La_{0.7}Sr_{0.3}MnO_3$(LSMO) and NiO were prepared by pulsed laser deposition using KrF excimer laser ($\lambda$=248 nm, pulse width 20 ns). Two single phase separate targets of (LSMO) and NiO were alternately used for deposition. Each target is hit with 14 laser shots alternatively in sequence with 3 Hz laser repetition rate for the duration of one hour to prepare the composite VAN structure in multi target deposition chamber. The composite thin films were simultaneously deposited on single (001) oriented crystalline substrate $LaAlO_3$ (LAO) and $SrTiO_3$ (STO) named as L14 an S14 respectively. Before the deposition, the substrates were ultrasonically cleaned with acetone bath and then methanol to remove any contamination or oil on the substrate surface. We deposited both the materials in 250 mTorr oxygen partial pressure with substrate temperature 750 ˚C and the energy flux at the target was kept at 2.0 J/cm$^2$ with 3 Hz repetition rate. After the deposition, the films were cooled in 0.5 b
ar with 5˚C/min cooling rate. From thickness measurements of such individual LSMO and NiO films, their effective thickness is estimated to be ~ 120 nm and 45 nm respectively, in the composite thin film L14 and S14. For comparison, we have also prepared epitaxial bilayer 50 *nm* NiO (at top) / 50 *nm* LSMO (at bottom) on STO (001) substrate at the same parameters and this bilayer hereafter is denoted as BL. For the structural characterization, the θ-2θ x-ray diffraction measurements were performed by Bruker D2 PHASER and reciprocal space mapping (RSM) are performed by D8 discover high resolution x-ray diffractometer. For the microstructure study, we have used field emission scanning electron microscope (FE-SEM) NOVA Nano SEM-450 by FEI. To study the emergence of SEB and CEB phenomenon magnetic measurements of the composite thin films were performed by 7 Tesla SQUID-vibrating sample magnetometer (SVSM; Quantum design, Inc., USA). To study the interfacial magnetic and electronic properties room temperature x-ray absorption spectroscopy (XAS) at Ni, Mn *L*-edge and O *K*-edge were performed at Indus-II, RRCAT, India, in total electron yield (TEY) mode. Temperature dependent x-ray magnetic circular dichroism (XMCD) measurements (up to lowest possible temperature 80 K) were performed at 2A MS undulator beamline under the 0.6 T magnetic field at Pohang accelerator laboratory (PAL) Korea respectively.

**Results and discussion:**

The *θ-2θ* x-ray diffraction (XRD) patterns of S14 and L14 are shown in Figs. 1(a) and 1(b) along with those of STO and LAO substrate. From the XRD patterns, it is noted that corresponding to NiO phase, only (002) reflection is observed on both substrates. Corresponding to LSMO phase (00ℓ) reflections are clearly observed on LAO substrate. However, these (00ℓ) reflections of LSMO are smothered under the (00ℓ) reflection of STO substrate because the out of plane lattice parameters of LSMO are stressed to match with STO through the vertical strain imparted by the immiscible NiO secondary phase in the composite thin films, as also discussed later in the manuscript.

FE-SEM micrographs shown in the Figs. 2 (a) and (b) for S14 and L14 samples respectively reveal their nano columnar and nano maze like VAN microstructures respectively. Self-assembled ultra-fine NiO nano-pillars with diameter of ~ 4-6 nm embedded in LSMO matrix is clearly seen in Fig. 2(a) for sample S14, whereas nano-maze like structure is observed in L14 sample [Fig. 2(b)]. Such different micro-structure in self-assembled nanocomposite thin films are reported earlier to arise due to the competition between surface energy, volume energy, and strain energy.[26] The side view of the L14 sample [Fig. 2(c)] reveals the columnar like growth of the LSMO/NiO composite films in both the samples suggesting their VAN structures. As discussed earlier, in S14 sample (00ℓ) reflections of LSMO are merged with the (00ℓ) reflection of STO substrate, which generally does not occur when only LSMO film is grown on STO substrate.[27] For instance, in the case of bilayer grown on STO, (002) reflection of LSMO is clearly visible as shown in the inset of Fig. 1(a). In S14 composite film, because of its contact with the surrounding LSMO, NiO (lattice parameter = 4.17 Å) will exert the out of plane tensile strain on LSMO (pseudocubic lattice parameter = 3.88 Å) while the underlying STO (3.905 Å) substrate will exert the in plane tensile strain. Thus, the vertical strain imparted by the NiO secondary phase along with the STO (3.905 Å) substrate induced in-plane strain, combinedly settles the unit cell lattice parameters of LSMO. However, in L14 sample, LSMO will experience in-plane compressive stress due to the LAO substrate (lattice parameter = 3.78 Å) leading to different microstructure than S14, as also evident from the Fig. 2. These observations clearly indicate towards the strong tuning of vertical strain between LSMO and NiO phase in composite VAN, which is very difficult to achieve in planer geometry.

To further understand the structural arrangement of LSMO and NiO on the LAO and STO substrates, reciprocal space mapping was recorded for the S14 and L14 samples. It is inferred from the RSM data that the LSMO fraction in the composite film is epitaxialy grown on LAO and STO substrates whereas epitaxy of NiO could not be observed. It is known that the surface energy of the grown materials and substrate play a key role in the wetting process and determines in such composite film as to which composition will act as host matrix for the other phase. In this combination of LSMO and NiO structures on STO substrate, LSMO will play the role of host matrix for the secondary phase of NiO.[28,29]

Magnetization as a function of the temperature of S14 and L14 films and LSMO/NiO bilayer are shown in Fig. 3 (a) in zero field-cooled cooling (ZFC) and field-cooled warming (FCW) cycles in the temperature range of 5-360 K under the in-plane 100 Oe applied magnetic field. Both VAN films and bilayer show clear paramagnetic to ferromagnetic transition, however, while the VAN films reveal magnetic transition of LSMO at $T_C$ ~225 K, the bilayer structure reveals $T_c$ at 300 K. The reduction of $T_c$ of the films with respect to the bulk LSMO could be [29]due to the microstructure along with presence of large in-plane as well as out of plane strain in the composite film induced due to underlying substrate and surrounding NiO, which can weaken the double exchange interaction via Mn-O-Mn bond angle and bond length, causing a rather broadened FM transition of LSMO.

A closer look at the M-T behavior of the composite films reveals anomaly at 180 K across which a sudden slope change in M-T is observed [lower inset of Fig. 3(a)]. This anomaly has been earlier attributed to the charge transfer between Ni and Mn ions at the interface of similar $La_{0.7}Ca_{0.3}MnO_3$ and NiO heterostructure.[30] As established later in the manuscript, it is found that the anomaly is not due to charge transfer but due to interaction between uncompensated moments at the surface/interface of LSMO and NiO phases.

In the M-T behaviour of the VAN samples S14 and L14, two notable features also appear: (i) cusp or peak ($T_p$) in the ZFC M-T cycle which disappears in FC cycle and (ii) bifurcation below the irreversibility $T_{irr}$.[31] For the sake of clarity, we shall focus on S14 sample onward for detail study, since L14 also displays qualitatively similar behaviour [inset Fig. 5(a)], though the features are more explicit in S14. Interestingly, ZFC magnetization versus applied field isotherm (M-H loop) measurement at 5 K for S14 sample reveals that the initial magnetization (virgin curve) lies outside the M (H) hysteresis envelope [see top inset of Fig. 3(a)]. This suggests toward the metastable magnetic state of the grown composite thin films which could be due to spin glass like behaviour.[32,33,34] It is noted that $T_P$ gradually shifts towards lower temperature with increasing measuring field (Fig. 3(b)) suggesting that the frozen disorder spin state is suppressed by strong DC magnetic field. According to spin glass mean field theory, Almeida-Thouless (AT) line predicts $T_p \propto H^{2/3}$ in H-T space which reveals the spin glass phase transition.[35] The inset of Fig. 3(b) shows the dependence of $T_p$ with field which fits well with AT line, supporting the spin glass behaviour and gives a zero field spin glass freezing temperature ($T_g$) ~ 106 K. These observations indicate towards the glass like metastable magnetic behavior.[36]

To further ascertain the spin glass behaviour of the film, we followed some established DC magnetization measurement protocols for such glassy systems viz. thermal remnant magnetizations (TRM), memory effect and rezuinuation.[37–39] TRM at different temperatures as shown in Fig. 4(a), at 50 K (where the spins are frozen), 80 K (below the $T_g$) and 140 K (above the $T_g$) under the 100 Oe applied fields. TRM behaviour is fitted by the stretched exponential function:

$$M(t) = M(0) \exp [-C (\omega t)^{1-n} / (1-n)] \ldots\ldots\ldots\ldots\ldots(1)$$

where ω is the relaxation frequency and C is exponential factor and *n* is the fitting parameter. From the fitting, *n* is found to be 0.71 and 0.80 at 50 K and 80 K respectively. These values of *n* are consistent with the previously observed interfacial spin glass state in such epitaxial bilayer such as $La_{0.7}Sr_{0.3}MnO_3/SrMnO_3$ and $La_{0.7}Sr_{0.3}MnO_3/LaNiO_3$.[25,40] The relaxation of the thermal remnant magnetization is most dominating near the $T_g$. Beyond $T_g$ at 140 K, as expected such relaxation is not observed.

To further divulge on the magnetic properties we performed ZFC relaxation memory experiments with well-designed DC magnetization protocols.[37] In the ZFC memory experiment the sample is cooled down to temperature $T_1$ (20 K) in zero field, and then 50 Oe field is applied and immediately magnetization [$M(T_1)$ is recorded as a function of time for *t* = 6000 seconds. After time t, the sample was quenched to a lower temperature $T_2$ (10 K) and $M(T_2)$ was recorded for time *t* seconds. Finally, the temperature is again brought to $T_1$ (20 K) and $M'(T_1)$ is recorded for time *t* seconds shown in Fig. 4(b). The relaxation in segment $M(T_2)$ at lower temperature is negligible as compared to segment $M'(T_1)$ at higher temperature. According to Sun et al.[37], for the spin glass like phase, the relaxation $M'(T_1)$ should be continuation of the relaxation at $M(T_1)$. The inset of Fig. 4(b) shows the continuation of relaxation at $M(T_1)$ and $M'(T_1)$ by neglecting the data points at $M(T_2)$. In Fig 4(c) we show the relaxation under the same protocol but with zero field in segment $M(T_2)$. In this protocol also the relaxation at $M(T_1)$ and $M'(T_1)$ are found in continuation without any apparent magnetization shift (inset of Fig. 4(c)) as expected for glass like dynamics. These results indicate that the state before temporary cooling (with or without constant field) is recovered when the temperature returns to $T_1$ (memory effect). According to hierarchical model, a small intermediate positive temperature cycling can destroy the memory effect.[37] We test this by introducing a small intermediate heating instead of cooling in the same protocols (Fig. 4(d)). It can be seen that the small intermittence heating fully reinitializes the relaxation and when temperature returns: $M(T)$ is not restored to the level before the temporary heating, i.e. memory effect is destroyed. The observation from TRM and ZFC memory experiments, further suggest the spin glass like phase in the composite film. Now we we discuss the exchange bias phenomena in these composite thin films.

The ZFC magnetization isotherm (M-H loop) of S14 sample at 5 K interestingly, appears asymmetric in shape and shifted towards the negative field axis (as shown in inset of Fig. 3(a)), which is characteristic of EB. Such type of EB in the ZFC M-H loop is termed as spontaneous exchange bias (SEB).[23] In the descending branch of the M-H loop, a step like feature is also observed, suggesting the step-wise magnetic reversal process in the system. This type of spontaneous exchange bias has been earlier observed in various systems mainly in super spin glass based interface in bulk, nano composite and superlattice.[32,39,34,41,42] To further investigate the SEB in detail and exclude the possibility of artifacts, we recorded the ZFC M-H loop at 5 K in two different ways (i) positive run (*p*-run) with sweeping the field as +1.5T→0T→ −1.5T→0T→+1.5T and (ii) negative run (*n*-run) with sweeping the field as −1.5T→0T→+1.5T→0T→ −1.5T [as shown in Fig. 5(a)]. The exactly opposite asymmetric ZFC M-H loops can be readily seen, confirming the intrinsic nature of the SEB effect of the studied composite structure. We shall like to emphasize here that the observed SEB is not due to minor loop effect since the M-H loop is well saturated at field value ~ 1T as shown in the inset of Fig. 5(c).[43]

The value of exchange bias from the hysteresis loop are calculated by using the formula $H_{EB} = |(-H_A + H_B)|/2$, where $H_A$ and $H_B$ are the magnitude of reverse and forward coercive fields. Thus the value of spontaneous exchange bias are found to be $H_{SEB}$ = 267± 5 Oe in *p* run and 270± 5 Oe in *n*-run is observed. In Fig. 5(b) dM/dH behavior shows the switching field behaviour of the sample in ascending and descending branches of ZFC M-H loop at 5 K. The two maxima in dM/dH in each branch can be associated with the respective soft (appearing at lower field value) and hard magnetic phases (at higher field value) leading to different magnetization reversal. Interestingly, the center of maxima corresponding to soft phase in ascending and descending cycles remain centered around zero suggesting this phase to remain unbiased, whereas the hard phase explicitly shows a clear bias with EB field = 260 ± 5 Oe along the field axis. These hard biased and soft unbiased M-H loops are de-convoluted by the equation (2),[44] as shown in Fig. 5(d).

$$M(H) = \sum_{i=1}^{2} \frac{2M_s^i}{\pi} \tan^{-1} \left| \frac{(H \pm H_c^i)}{H_c^i} \tan\left(\frac{\pi S^i}{2}\right) \right| \dots\dots\dots\dots\dots\dots (2)$$

Where $H_C$ is the coercivity, $M_S$ is the saturation magnetization, and S is the ratio of remnant to saturation magnetization ($M_R/M_S$).

Generally, the individual LSMO epitaxial film reveals unbiased soft ferromagnetic phase.[27] Whereas, in the studied self-assembled LSMO:NiO composite films, various competing exchange interactions occur at the interface among $Ni^{+2}$, $Mn^{+3}$ and $Mn^{+4}$ ions, which cause magnetically disordered state at interface. Thus it is inferred that the unbiased soft phase is due to reversal of FM LSMO which is not interacting with NiO, whereas biased hard phase is arising from the reversal of FM LSMO coupled with NiO through magnetically disordered interface. Superposition of these two contributions leads to asymmetric loop shape in ZFC M-H.[45] Qualitatively similar behaviour is observed in L14 sample also. Magnetic hysteresis loop recorded at 5 K after 1.5T (beyond the technical saturation) field cooled cooling (FCC) from 360 K, clearly shows [Fig. 5(c)] the shift along the field axis without step feature indicating the existence of conventional exchange bias (CEB) also. In the SEB, the enhanced coercivity and asymmetry in magnetization isotherm with respect to CEB isotherm at 5 K [Fig. 5(c)] suggests that the pinning of moments is prominent at magnetically disordered interface[46,47] between LSMO and NiO rather than conventional ordered FM/AFM interface. In CEB, the interface magnetic disorder state is supressed during the FCC process causing the reduction in EB field value.

To explore the origin of SEB it is crucial to study the dynamics of spin structure of AFM at the interface through training effect. The training effect measurements (consecutive repeating M-H cycles '*n*' times) performed at 5 K for both SEB and CEB [shown in Figs. 6(a) and 6(b) respectively] represent the process of spin rearrangement at the interface. In SEB, we observe a strong training effect. The major asymmetry of the hysteresis loop starts disappearing with increasing the loop cycling and the asymmetric loop is transformed in to the symmetric one at higher *n* value (5) and the step feature in M-H vanishes [see Fig. 6(a)]. This further confirms that the observed SEB is due to magnetically disordered interface rather than the ordered FM-AFM interface. In a theoretical study, Usadel et al.[48] reported that in the SG/FM interface, the interface exchange field acts like the interface magnetization of the AFM layer in conventional FM/AFM systems. A monotonous decrease in $H_{EB}$ and $H_C$ [$(H_A + H_B)/2$] with increasing '*n*' is observed in SEB as well as CEB due to interfacial spin rearrangement, which takes place in each cycle and modify the coercivity and exchange bias filed [see Figs. 6(c) and 6(d)].

It should be noted that the empirical law for the exchange bias in ordered AFM spin arrangement[49,50] at the FM/AFM interface is modelled as $H_{EB}^n = H_{EB}^\infty + kn^{(-1/2)}$. This empirical law does not fit well with our experimental data either in SEB or CEB, for example CEB, as shown by dashed line in Fig. 6(e) in CEB, which was the best possible fit to the data with $H_{EB}^\infty$ = 84 ± 7 Oe and $k$ = 118 ± 11 Oe. Earlier spin glass model of the exchange bias was proposed[39,51,52] which considered two different types of AFM uncompensated spins named as frozen-in/irreversible (*f*) AF spins and rotatable/reversible (*r*) spins rigidly exchange coupled to FM layers. During reversing the field, the rotatable spins follow the FM layer rotation whereas, frozen-in AF spins remain unchanged as illustrated in the schematics in Fig. 7. Thus both contribute distinctively with different relaxation rates during spin rearrangements in the hysteresis cycling. Considering this scenario we have simulated the relaxation of exchange bias as function of '*n*' by the equation (3):[51,39]

$$H_{EB}^n = H_{EB}^\infty + A_f exp^{(-n/P_f)} + A_r exp^{(-n/P_r)} \ldots\ldots\ldots\ldots\ldots (3)$$

Here $H_{EB}^n$ is the exchange bias of the $n^{th}$ hysteresis loop, $A_f$ (have dimension of magnetic field) and $P_f$ (dimensionless parameters related to the relaxation process) are the parameter related to the change in frozen spins and $A_r$ and $P_r$ are the evolving parameters of the interfacial magnetic frustration at FM/AFM interface. The equation (3) fits the experimental data perfectly well [solid line in Fig. 6 (e)] with parameters $H_{EB}^\infty$ = 133.4 ± 0.4, $A_f$ = 205.0 ± 7.0, $P_f$ = 0.85 ± 0.03, $A_r$ = 14.2 ± 1.0 and $P_r$ = 3.1 ± 0.27. The ratio $P_r/P_f$ = ~ 4 indicates that the frozen spins relax or rearrange with 4 times slower as compared to rotatable spin at the interface. This suggests that in the M-H loop protocols, when magnetic field is switched from positive to negative direction, moments at magnetically disordered interface flip towards the negative H direction (see Fig. 7), while the remaining moments are still pointed towards the positive H direction and rearranging rather slow, causing novel asymmetry in the M-H loop with exchange bias.

To probe the thermal evolution of EB, we have recorded zero field cooled and 1.5 T field cooled (cooling from 360 K) M-H isotherm as function of temperature, shown in Figs. 8 (a) and (b). The temperature dependence of $H_{EB}$ and $H_C$ are fitted by equation (4).

$$H_{EB}(T) = H_{EB}^0 exp^{(-T/T_1)}, \ H_C(T) = H_C^0 exp^{(-T/T_1)} \ \ldots\ldots\ldots\ldots\ldots\ldots (4)$$

Where $H_{EB}^0$ and $H_C^0$ are the extrapolation of the $H_{EB}$ and $H_C$ at the absolute zero temperature, $T_1$ and $T_2$ are constants. Both the SEB and CEB fields decay exponentially with temperature and follow equation (4). This observation is consistent with previous reports for several such systems showing spin glass mediated EB effect viz. LSMO/SrMnO3 bilayer,[25] La$_{1-x}$Ca$_x$MnO$_3$ (x= 0.33, 0.4, 0.48)/La$_{1-y}$Ca$_y$MnO$_3$ (y= 0.52, 0.67, 0.75) FM/AFM superlattices.[53] It can be seen that CEB ceases beyond 55 K, while SEB appears below 35 K only. These temperature values (generally designated as blocking temperature $T_B$ for such systems) are much lower than the Neel temperature of NiO (520 K). Therefore, it is clear that beyond $T_B$ the AFM order or spin glass phase is not stable enough to provide the unidirectional anisotropy to produce EB. It is to be noted that the $T_B$ is much below the spin glass freezing temperature $T_g$ indicating that interfacial competing magnetic order plays a very crucial role to establish and control such unusual exchange bias effect.

In order to divulge the microscopic origin of such unusual magnetic behaviour and henceforth the exchange bias phenomena, we examine the valence state of Mn, Ni and the relative alignment of their magnetic moments at the interface of the LSMO and NiO by performing the elemental specific XAS and XMCD measurements in TEY mode. Room temperature Mn $L$-edge and Ni $L$-edge XAS spectra of S14 and L14 samples are shown in Figs. 9(a) and 9(b), along with reference samples LSMO and NiO respectively. XAS Mn $L$-edge features in the composite films are similar to the LSMO film, suggesting the expected mixed valence state of Mn$^{3+/4+}$. It should be noted that with Ni $L_3$-edge, it is very difficult to quantify the Ni oxidization state, due to the overlapping tail of La $M_4$-edge with Ni $L_3$-edge at ~ 853 eV. However, analysing Ni $L_2$-edge is decisive in divulging the ionic state of Ni.[54] As shown in Fig. 9(b), we observe a clear splitting in Ni $L_2$-edge, revealing the Ni$^{2+}$ valance state of Ni ion, as expected for NiO, under the octahedral environment.[55] To further substantiate the 2+ state of Ni, we simulated the $L_2$-edge spectra for Ni$^{3+}$ and Ni$^{2+}$ states using CTM4XAS code[56] and compare them with the experimental data as shown in the inset of Fig. 9(b). This explicitly confirms the Ni$^{2+}$ state in the composite films. Similar spectra are also observed in L14 sample.

To get an insight of the magnetic anomaly observed below 180 K, as discussed earlier, we recorded XAS and XMCD spectra at lower temperature values such as 180, 100 and 80 K. In Figs. 9(c) and (d) we show Mn $L$-edge and Ni-$L$-edge spectra recorded at 100 K. It is clear that the features of Ni $L_2$-edge and Mn $L$-edge remain unchanged from the spectra recorded at 300K. In previous studies[30] it was argued that the magnetic anomaly observed in similar composite structures comprising of $La_{0.7}Ca_{0.3}MnO_3$ and NiO at lower temperature (< 90K) arises due to the charge transfer mechanism between the $Ni^{2+}$ and $Mn^{3+/4+}$. However, we do not observe any signature of charge transfer even down to 80 K. This indicates that the observed anomaly may arise due to onset of magnetic interaction between $Mn^{3+/4+}$ and $Ni^{2+}$ ions, rather than charge transfer mechanism.

The XMCD spectra at the Mn $L$-edge and Ni $L$-edge recorded under 0.6 T magnetic field at the selected temperature values (at 320, 180 and 100 K) are shown in Fig. 9(e). The temperature values are chosen so as to probe the nature of interaction between Ni and Mn ions, when (i) LSMO fraction of the composite film reveals paramagnetic behaviour (at 320 K), (ii) LSMO is ferromagnetic (at 80 K) and (iii) magnetic anomaly temperature (at 180 K). It is worth mentioning here that these composite films saturate at ~0.3 Tesla. Ni $L$-edge does not reveal any XMCD signal at 320 K [Fig. 9(f)], as expected from antiferromagnetic NiO with Neel temperature 520 K. Interestingly, XMCD signal at Mn $L$-edge can be spotted at 320 K, albeit small, despite the M-T data revealing FM-PM transition at much lower temperature value than generally observed for bulk LSMO as discussed earlier. It should be noted here that the presented XMCD is recorded at 6000 Oe, whereas the M-T data shown in Fig. 3(a) is recorded at much lower magnetic field of 100 Oe. However, when the M-T data is recorded at higher measuring field values from 50 Oe to 2000 Oe, as shown in Fig. 3(b) a systematic rise in FM-PM transition can be easily noticed. Therefore, observation of feeble XMCD signal can be attributed to the FM interaction due to Zener double exchange interaction between $Mn^{3+}$ and $Mn^{4+}$ in LSMO. At 180 K, Mn $L$-edge explicitly shows ferromagnetic polarization. Interestingly at 180 K which coincides with the magnetic anomaly temperature, $Ni^{2+}$ shows XMCD signal suggesting ordering of Ni moments also,[57] as shown in Fig. 9(f). It is important to note that the XMCD signal [Figs. 9(e) & (f)] is largely negative at the $L_3$-edge for both Mn and Ni which is increasing with cooling, suggesting that the $Ni^{2+}$ and $Mn^{3+/4+}$ ions are ferromagnetically coupled at the interface.[58]

NiO is known to be antiferromagnetic with Neel temperature ($T_N$) 520 K, so its XMCD signal is not expected, as observed in XMCD spectra at 320 K. However, the signature of Ni moment ferromagnetically aligned with the Mn ions in the composite film at 180 K in the XMCD signal points out towards the stronger ferromagnetic interaction between $Mn^{3+/4+}$ and $Ni^{+2}$ at the LSMO-NiO interface than the antiferromagnetic Ni-Ni coupling in NiO. This also insinuates towards a possible proximity effect driven ordering of uncompensated Ni moments in AF NiO at the interface. During the zero field cooling process, such Ni-Mn ferromagnetic interaction together with the FM LSMO and AFM NiO at the interface will give rise to frustrated magnetic phase at the AFM-FM interface leading to SEB below the spin glass transition temperature. From these discussions it is transpired that the glass like disorder phase at the interface is an important ingredient to establish the zero field exchange bias in the system.

## Conclusions:

In conclusion nano composite thin films of LSMO:NiO have been grown by pulsed laser deposition. We have presented the systematic experimental investigation of the unusual exchange bias in the nanocomposite thin films. The DC magnetization measurements show the spin glass like transition around 106 K in the S 14 sample. Below this spin glass freezing temperature, we have found the unusual zero field cooled spontaneous exchange bias (SEB) with novel asymmetry in the hysteresis loop, along with conventional exchange bias (CEB) effect. The blocking temperature for SEB is around 35 K whereas for CEB it is 55K. The exchange bias relaxations in training effect is found to be disorder mediated, arising from contributions due to different field response rates of frozen-in and rotatable spins at the interface. From temperature dependent magnetization and XMCD measurements we conclude that SEB is mainly originating due to unstable interfacial AFM (NiO) spin alignment of the disordered glass like phase which is ferromagnetically coupled with the FM (LSMO).

## Acknowledgments:

The authors are thankful to Dr. V. Raghavendra Reddy for RSM measurements. We are also thankful to Dr. Dileep Kumar for fruitful discussions. Authors also acknowledge Mr. Rakesh Kumar Sah for help during XAS measurements.

**Figure Captions:**

**Figure 1.** (a) *θ-2θ* X-ray diffraction patterns of composite thin film S14 grown on single crystalline STO (001), L14 grown on single crystalline LAO (001) substrate along with bare substrate STO and LAO. Top inset shows the XRD of LSMO/NiO bilayer grown on STO substrate and close view of (002) reflection of S14 sample.

**Figure 2.** FE-SEM top micrograph of (a) S14 (nano columnar), (b) L14 (nano maze) LSMO:NiO composite thin films and (c) cross-sectional view of L14 sample. Reciprocal space maping of L14 sample is shown around symmetric (002) Bragg reflection (d) and asymmetric (-103) Bragg reflection (e) of LAO.

**Figure 3.** (a) Magnetization versus temperature (M-T) behavior of S14, L14 composite and LSMO-NiO bilayer thin films with zero field-cooled cooling (ZFC) and 100 Oe field-cooled warming (FCW). Upper inset shows the ZFC asymmetric M-H of S14 recorded at 5 K, lower inset shows close view of anomaly near 180 K. (b) M-T of the S14 composite thin film under different magnetic fields with ZFC and FCC. Inset shows the $T_p$ vs $H^{2/3}$ behaviour fitted with AT line.

**Figure 4.** (a) Time dependence of thermal remnant magnetization after 100 Oe field cooling from 340 K to desired temperature. Magnetization relaxation in intermediate negative temperature cycling (b) in ZFC method without field changing (c) in ZFC method with field change. (d) Magnetization relaxation in intermediate positive temperature cycling in ZFC method without field changing.

**Figure 5.** (a) Magnetic hysteresis loops of S14 sample recorded at 5 K in two different field sweeping after zero field cooled in *p*-run and *n*-run, inset shows the ZFC M-H of L14 at 5K (b) dM/dH behaviour of the sample in ascending and descending branches of ZFC M-H loop at 5 K (c) 1.5 Tesla FC Conventional and ZFC spontaneous exchange bias isotherms at 5K, inset shows the saturated M-H of S14 at 5K (d) deconvoluted soft (unbiased) and hard (biased) phase from zero field cooled M-H at 5K.

**Figure 6**. Impact of training effect at 5 K on 1$^{st}$, 2$^{nd}$ and 5$^{th}$ loop in (a) SEB, and (b) 1.5 T FC (cooled from 360 K) CEB. Reduction in coercivity and exchange bias field as the function of repeating hysteresis loop index *n* is shown in (c) for SEB and (d) for CEB. (e) Solid star represents the experimental data, dashed line represent the 1/√n function fit and solid line show the fitted model by equation (3).

**Figure 7.** Simplified schematic representation of magnetization reversal of frozen-in and rotatable spins at the interface during the training.

**Figure 8.** M-H isotherms of sample S14 recorded at different temperatures after (a) zero field cooling from 360 K, (b) 1.5 T field cooled cooling from 360 K. The temperature dependence of $H_C/H_{SEB}$ and $H_C/H_{CEB}$ for S14 are shown in (c) and (d) respectively, solid lines are the fittings of experimental data to equation. (4). Arrow mark the blocking temperature ($T_B$).

**Figure 9.** Room temperature x-ray absorption spectra of L14 and S14 collected at (a) Mn $L_{2,3}$-edge and (b) La $M_4$ and Ni $L_{2,3}$-edge along with LSMO, NiO reference samples. Left and right circularly polarized light XAS of S 14 are shown for Mn $L$-edge in (c) and La $M_4$-Ni $L_{2,3}$-edge in (d). The respective XMCD spectra with temperature are vertically shifted for clarity and shown in (e) & (f).

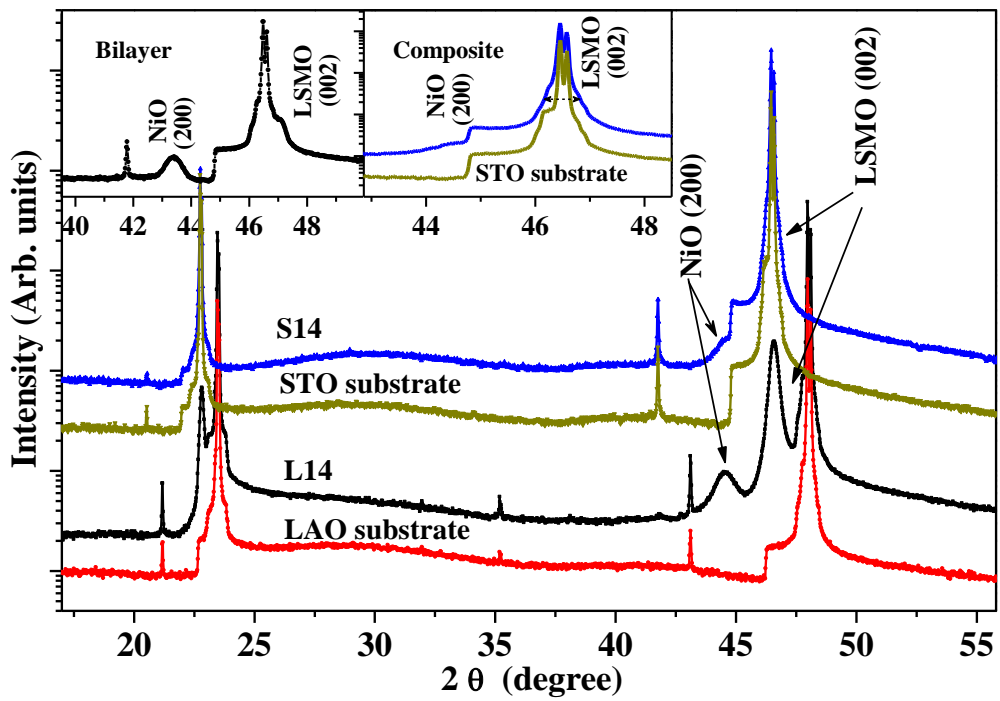

**Figure 1**

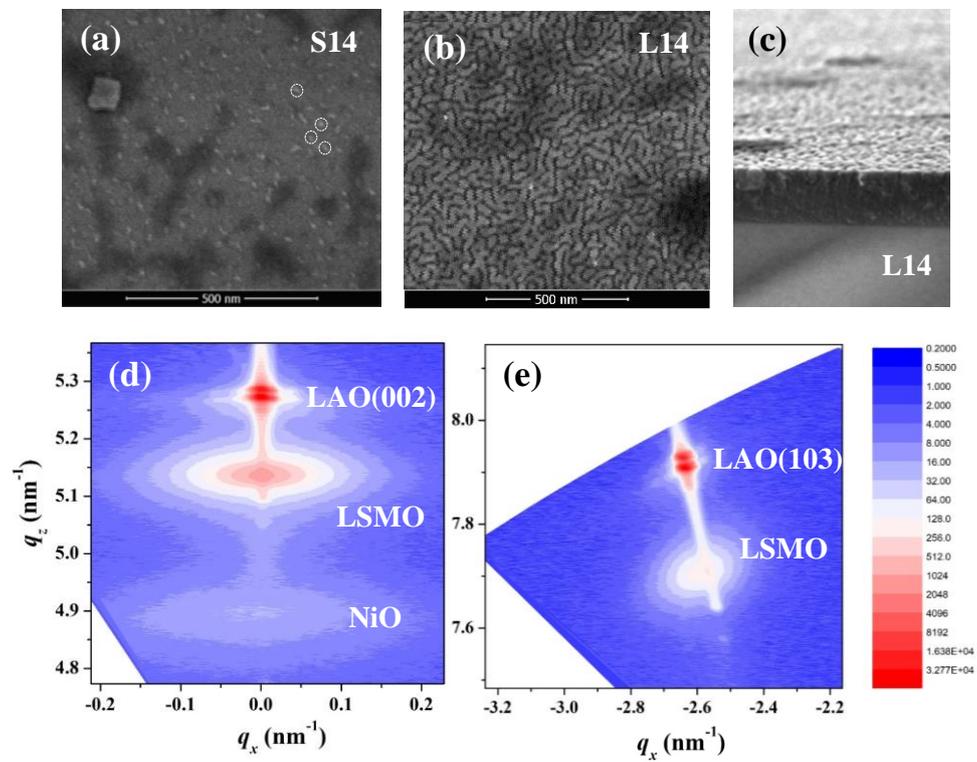

**Figure 2**

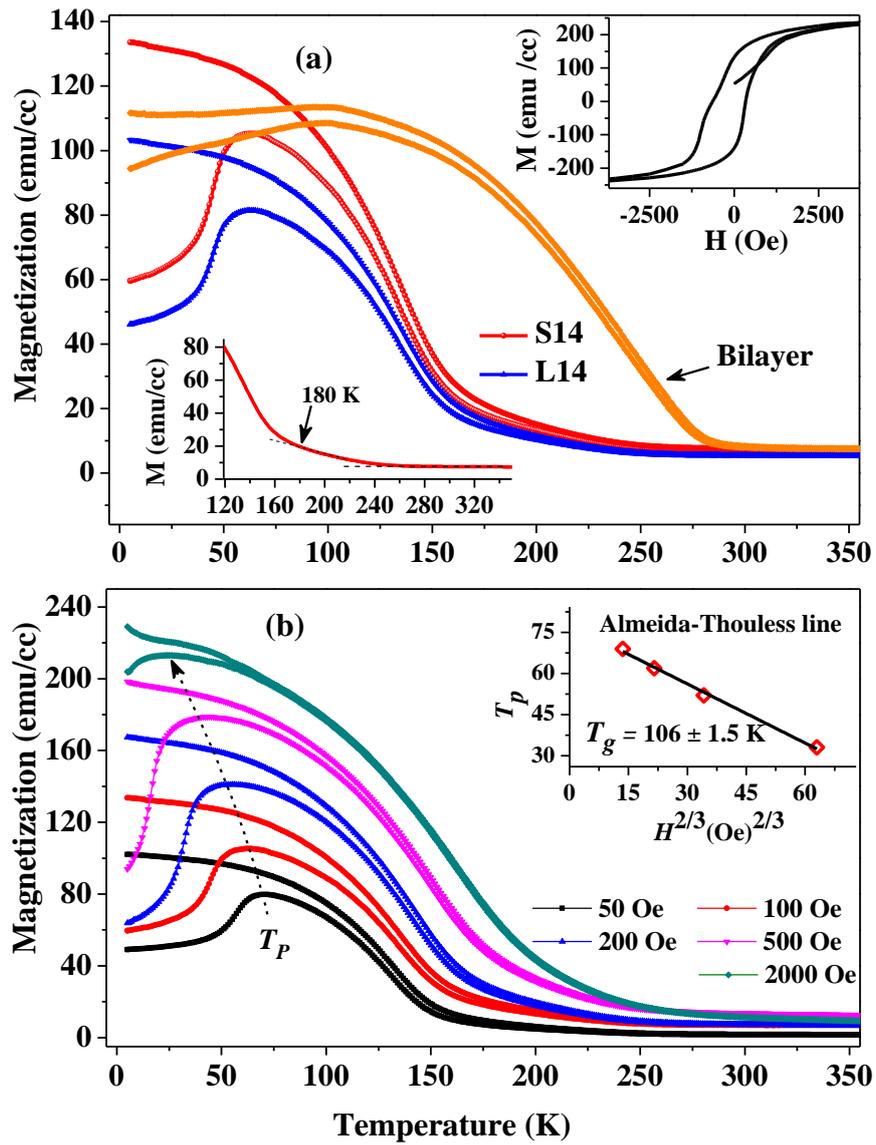

Figure 3

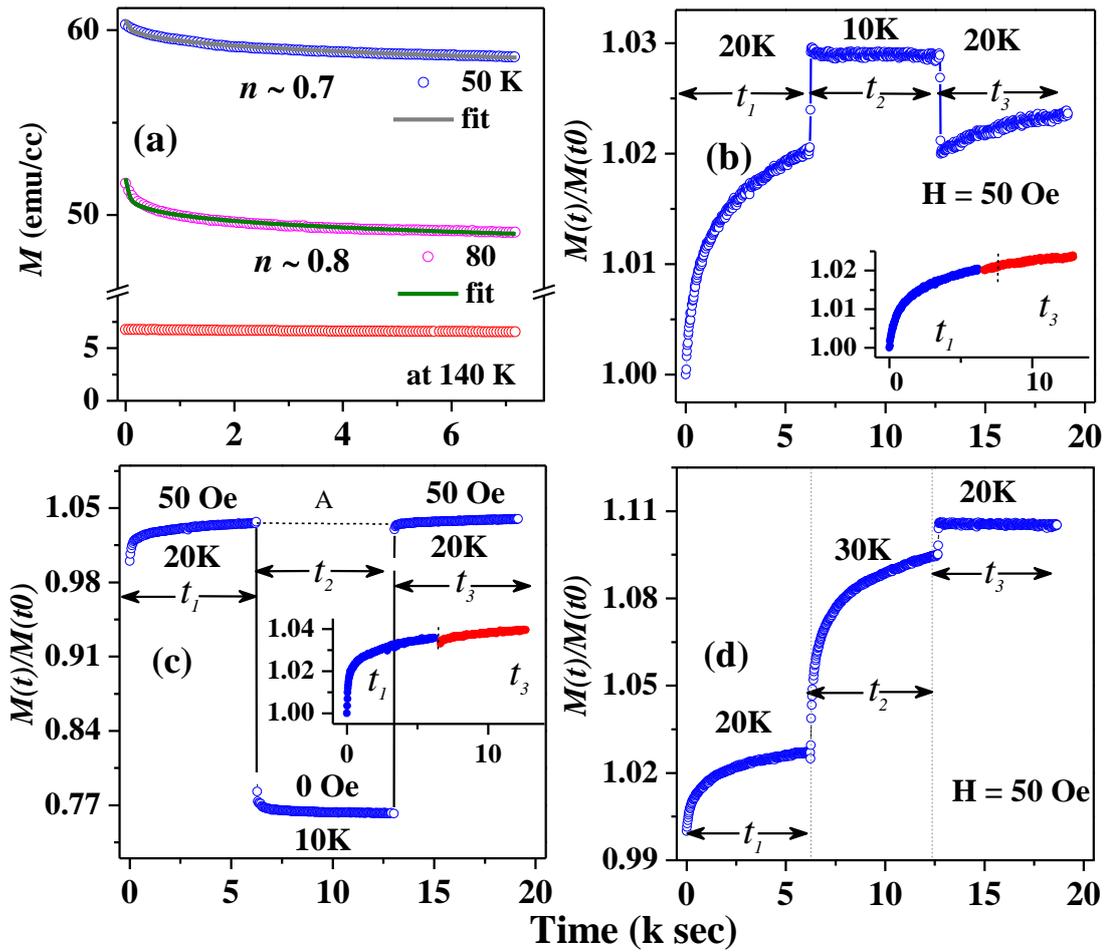

**Figure 4**

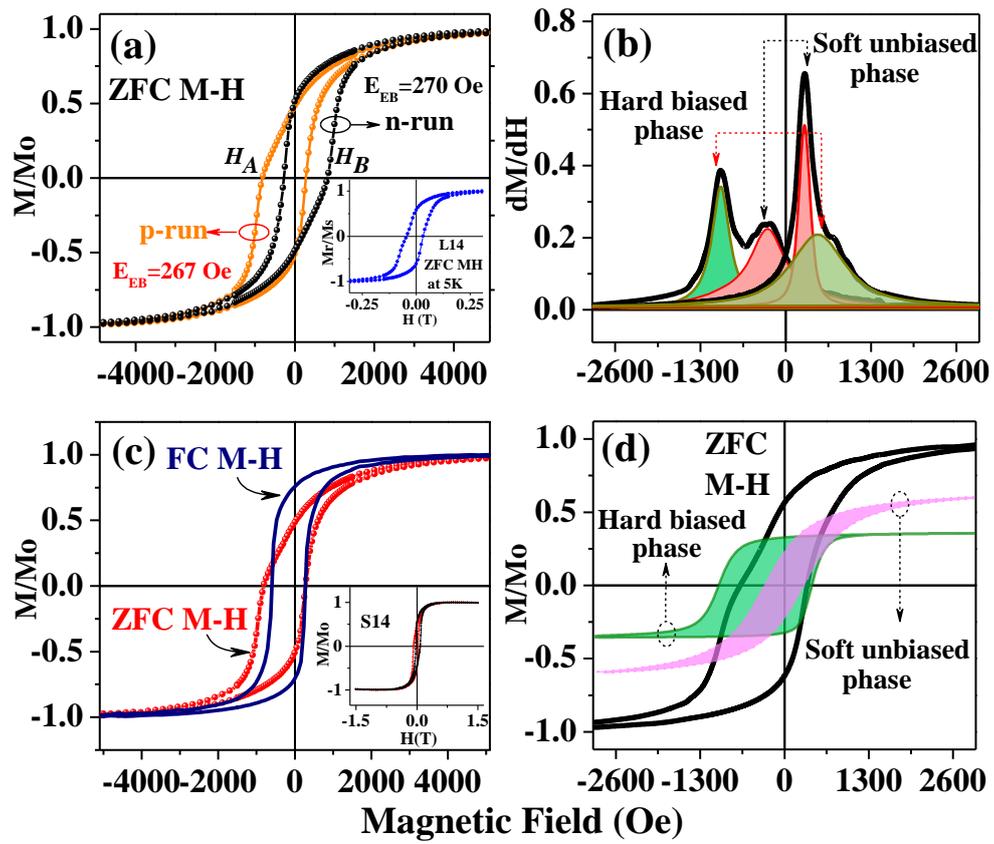

Figure 5

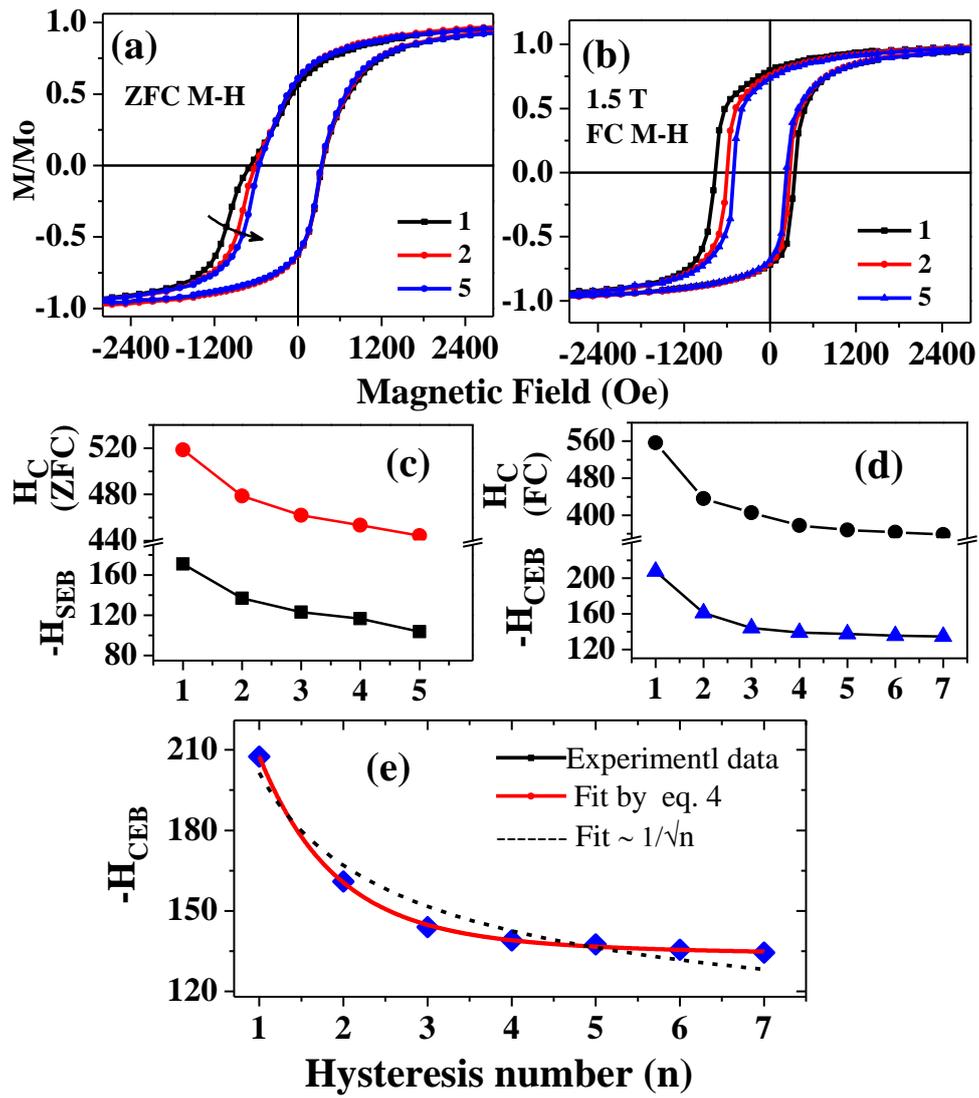

Figure 6

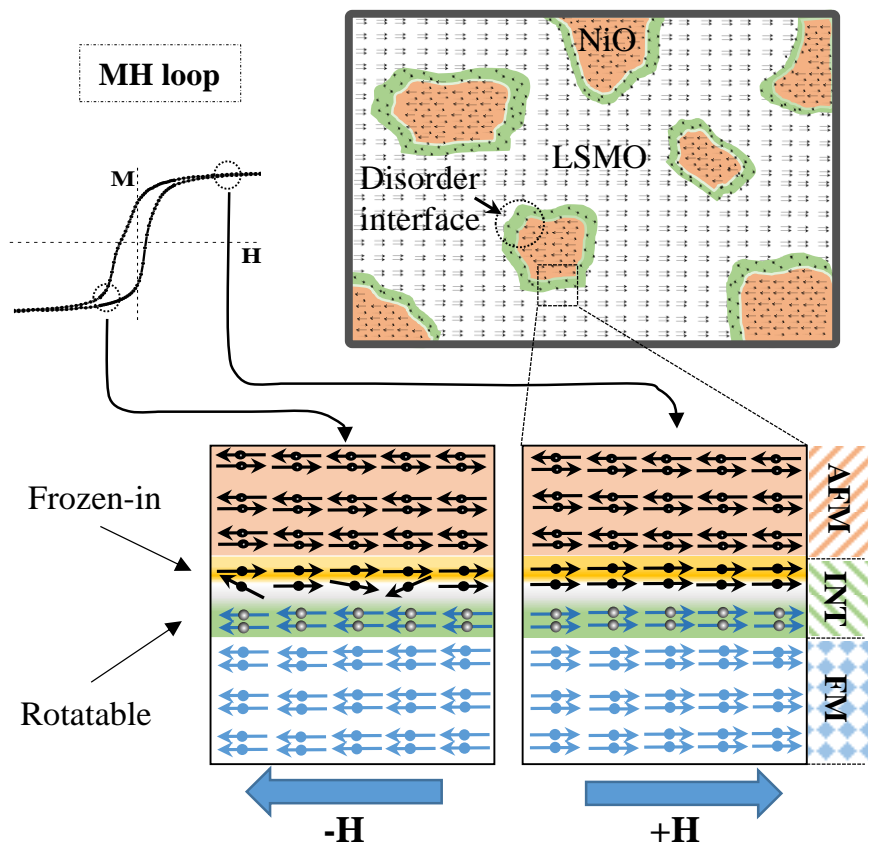

**Figure 7**

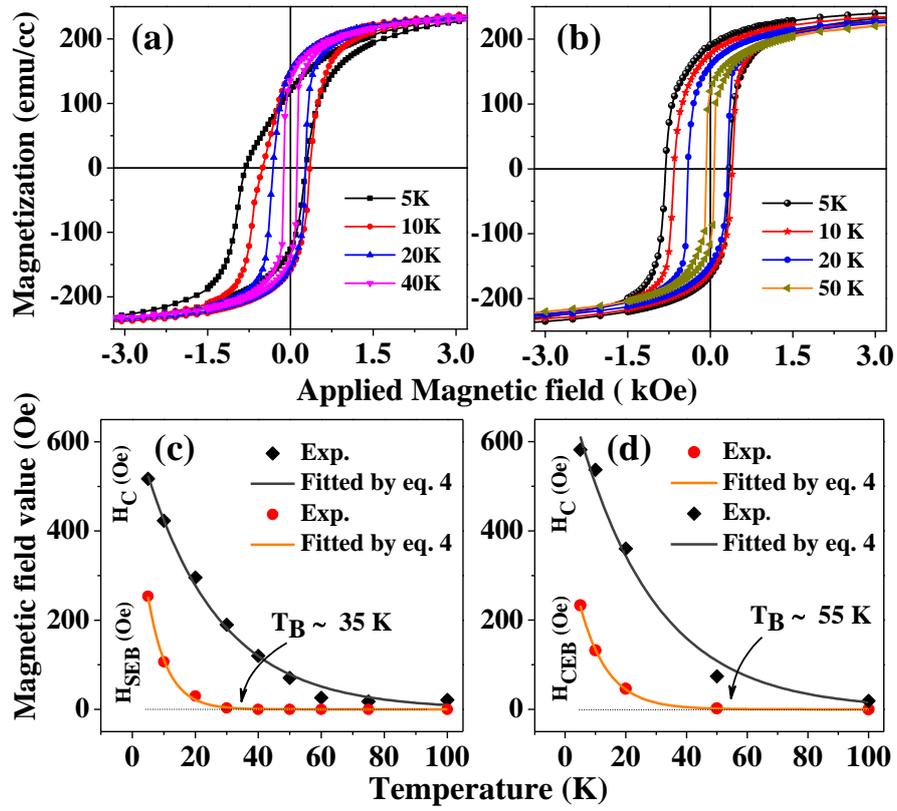

Figure 8

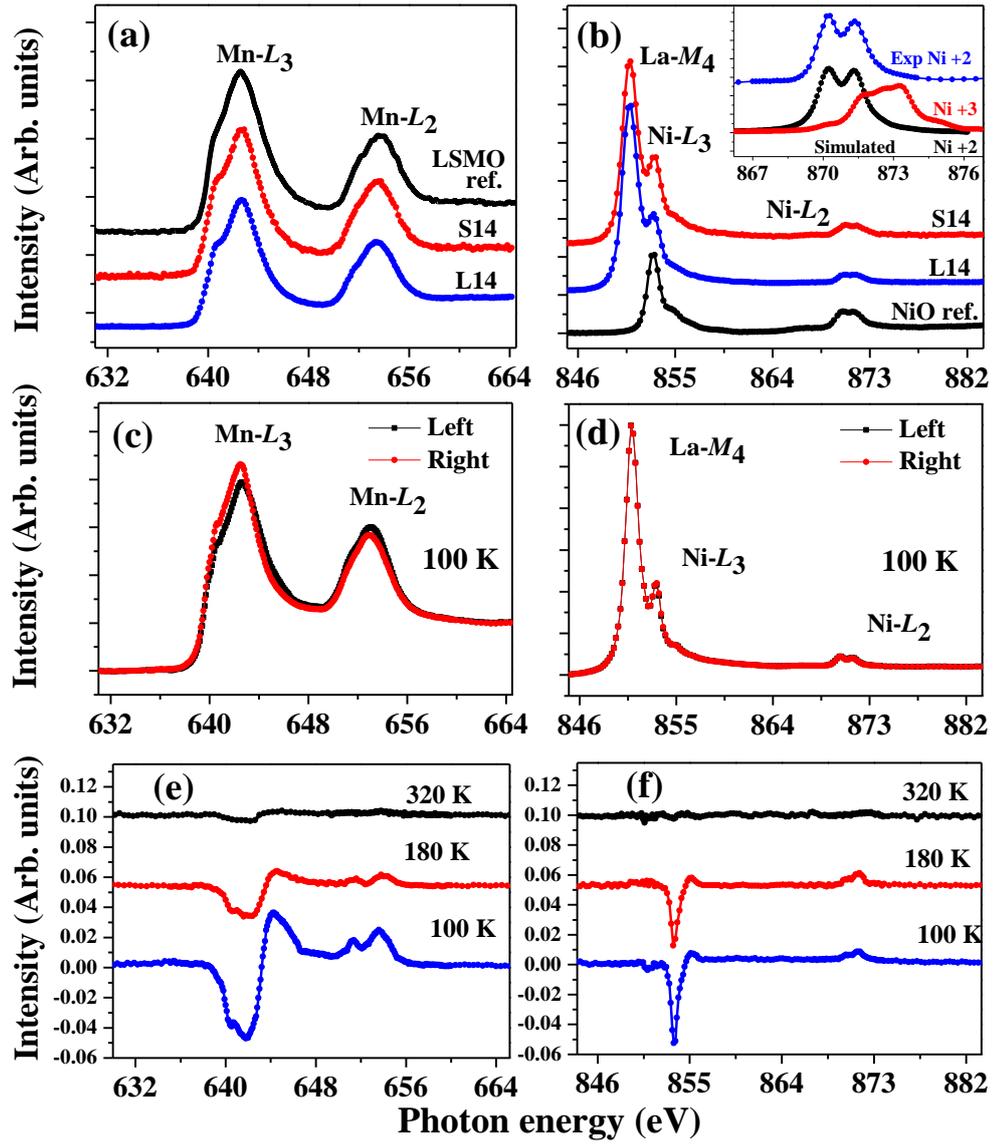

**Figure 9**